# Melting Domain Size and Recrystallization Dynamics of Ice Revealed by Time-Resolved X-ray Scattering

Cheolhee Yang[1], Marjorie Ladd-Parada[2], Kyeongmin Nam[1], Sangmin Jeong[1], Seonju You[1], Alexander Späh[2], Harshad Pathak[2], Tobias Eklund[2], Thomas J. Lane[3], Jae Hyuk Lee[4], Intae Eom[4], Minseok Kim[4], Katrin Amann-Winkel[2], Fivos Perakis[2], Anders Nilsson[2], and Kyung Hwan Kim[1,*]

[1]Department of Chemistry, Pohang University of Science and Technology (POSTECH), Pohang, Gyeongbuk 37673, Republic of Korea.

[2]Department of Physics, AlbaNova University Center, Stockholm University, SE-106 91 Stockholm, Sweden.

[3]SLAC National Accelerator Laboratory, 2575 Sand Hill Road, 'Menlo Park, CA 94025, USA.

[4]Pohang Accelerator Laboratory, Pohang, Gyeongbuk 37673, Republic of Korea.

*Corresponding author. Email: kyunghwankim@postech.ac.kr

**Abstract**

The phase transition between water and ice is ubiquitous and one of the most important phenomena in nature. Here, we performed time-resolved x-ray scattering experiments capturing the melting and recrystallization dynamics of ice. The ultrafast heating of ice I is induced by an IR laser pulse and probed with an intense x-ray pulse, which provided us with direct structural information on different length scales. From the wide-angle x-ray scattering (WAXS) patterns, the molten fraction, as well as the corresponding temperature at each delay, were determined. The small-angle x-ray scattering (SAXS) patterns, together with the information extracted from the WAXS analysis, provided the time-dependent change of the size and the number of the liquid domains. The results show partial melting (~13 %) and superheating of ice occurring at around 20 ns. After 100 ns, the average size of the liquid domains grows from about 2.5 nm to 4.5 nm by the coalescence of approximately six adjacent domains. Subsequently, we capture the recrystallization of the liquid domains, which occurs on microsecond timescales due to the cooling by heat dissipation and results to a decrease of the average liquid domain size.

The melting of ice and the crystallization of liquid water are amongst the most common and important phenomena in nature and everyday life[1-11]. Thus, it is crucial to understand the fundamental mechanisms of the phase transition between ice and water. The melting of ice has been investigated intensively by using time-resolved optical spectroscopies[12-16] and computer simulations[17-23]. However, the detailed mechanisms of the melting and recrystallization of ice in terms of the structural dynamics are not fully elucidated due to the lack of direct length scale sensitivity of the spectroscopic methods, and the limited size of the system in computer simulations.

In a study using time-resolved IR spectroscopy, a rapid temperature increase (T-jump), induced by an intense IR pump pulse, was previously utilized to investigate the homogeneous melting of bulk ice. In this study, experimental evidence of the superheating of ice on sub-ps to hundreds of ps timescales was observed[12]. More recently, using a similar approach with a wider time window, it was observed that the superheating and homogeneous melting of ice are temporally overlapped and only a partial melting of ice occurs within the first 25 ns, even when excess energy is provided[15]. Even though these spectroscopic studies have provided insight on the dynamics of melting and superheating of ice, more direct structural information, such as the number and size of the molten domains and how they change over time, is necessary to fully understand the mechanism of the phase transition between ice and liquid water. Moreover, previous studies have not elucidated the recrystallization process (crystallisation post-melting). An experiment utilizing time-resolved Mie scattering spectroscopy, with a millisecond (ms) time window, probed the time-dependent size of the molten domain and suggested a model for the melting and recrystallization dynamics. However, the direct estimation of the size was only possible at very late delays (after 0.3 ms)[16].

An x-ray probe can be the optimal tool for providing structural information[24-27]. By combining a rapid T-jump, induced by an IR laser pulse, with an intense x-ray pulse from free-electron lasers (FELs) as a probe (x-ray scattering), the detailed structural dynamics of the melting and recrystallization of ice can be elucidated. Furthermore, in x-ray scattering experiments, structural information with different length scales can be obtained depending on the observed scattering angle. While the wide-angle x-ray scattering (WAXS) is sensitive to the molecular structure and the local structure between adjacent molecules, small-angle x-ray scattering (SAXS) is sensitive to the size and shape of much larger domains. By combining the results extracted from SAXS and WAXS after the T-jump, a more comprehensive picture on the melting and recrystallization dynamics of ice with direct structural information can be elucidated.

In this study, we measured the time-resolved wide and small angle x-ray scattering pattern of ice I at various time delays before and after the T-jump to follow melting and recrystallization dynamics of ice at the XSS beamline of PAL-XFEL[28,29]. The experimental setup is shown schematically in Fig. 1(a). The detailed experimental procedure and analysis are described in a previous paper[26] and in the Method section. Briefly, an ice I sample with a thickness of ~84 μm was prepared on a Cu-grid and loaded on custom-made sample holder connected to a cryostat. The sample was then placed inside a vacuum chamber with a pressure of less than 0.1 Pa. Femtosecond IR laser pulses at 2 μm wavelength with a pulse energy of 250 μJ/pulse were focused to a spot size of 70 μm (FWHM) and were used to induce the ultrafast T-jump of the ice sample. Femtosecond x-ray pulses generated by PAL-XFEL were used to measure scattering patterns of the sample before and after the T-jump. The momentum transfer $q$-range of 0.1–3.0 Å$^{-1}$ was covered by a large area CCD detector (MX225-HS, Rayonix). To probe a fresh spot for each measurement, the sample holder was moved to an unexposed position prior to each pump-probe measurement. We note that our ice sample is mounted without using any windows and thus is directly exposed to the pump and probe pulses. Our setup provides the following advantages: (1) There is no background scattering from any window materials, (2) there is no extra cooling process that can be induced by the heat conduction between the sample and the windows, and (3) the total volume of the sample is not constrained, and it allows the sample to change in volume.

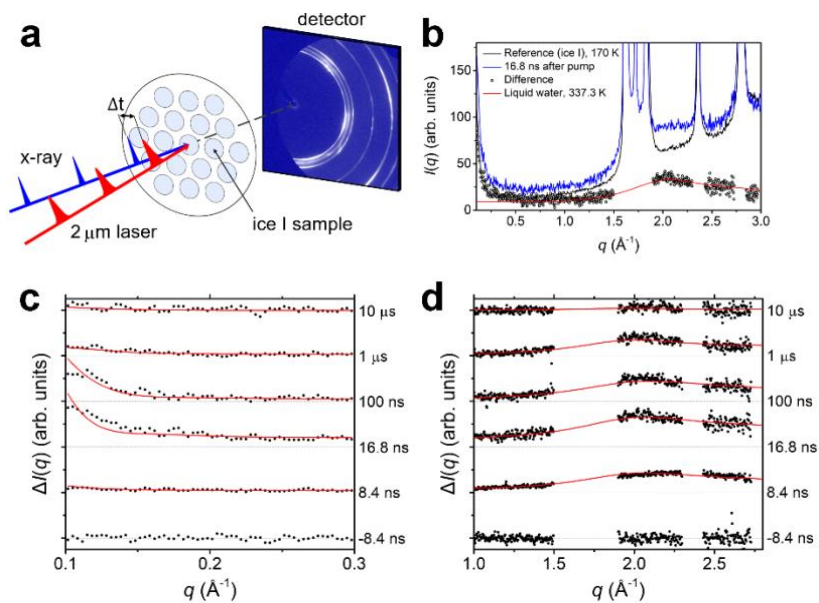

**Fig. 1.** (a) Schematic diagram of the experimental setup. Femtosecond IR laser pulses were used to induce an ultrafast T-jump of the ice sample. Femtosecond x-ray pulses were used to measure scattering patterns of the sample before and after the T-jump. (b) The angularly integrated x-ray scattering patterns before (black line) and after (blue line) the T-jump and the difference between them (black circles) are shown together. The difference scattering is compared with the scattering pattern of liquid water measured at 337.3 K (red line) taken from a previous study[30]. (c, d) The difference scattering intensities (black circles) in the (c) SAXS and (d) WAXS regions at time delays from -8.4 ns to 10 μs. The Bragg peaks of ice I are removed for clarity. The fitting results (red line) are shown together.

The experimental results of the IR laser-induced melting of ice I are summarized in Fig. 1. As an example, x-ray scattering intensities before the IR pump and 16.8 ns after the pump and the difference between them are shown together in Fig 1(b). The difference shows the generation of diffuse scattering, comparable to the scattering pattern of liquid water measured at 337.3 K (red line, taken from ref[30]), clearly indicating that the ultrafast T-jump causes the melting of ice I. We note that for clarity, the $q$ range that is affected by the Bragg peak of ice I is omitted in the difference pattern.

The difference scattering intensities in the SAXS and WAXS regions at various time delays from -8.4 ns to 10 μs are shown in Fig. 1(c) and Fig. 1(d), respectively. The difference intensities, both in the SAXS and WAXS regions, rise within tens of ns and decay at later delays. The rise in intensity can be associated to the melting of ice, whilst the later decay is indicative of the recrystallization process. Interestingly, the timescales of the decay of SAXS and WAXS are different as shown in Fig 2(a). The decay in the SAXS region is much faster than that in the WAXS region. Given that both regions provide structural information at different length scales, this discrepancy is the key to understanding the comprehensive picture on the structural dynamics of melting and recrystallization of ice.

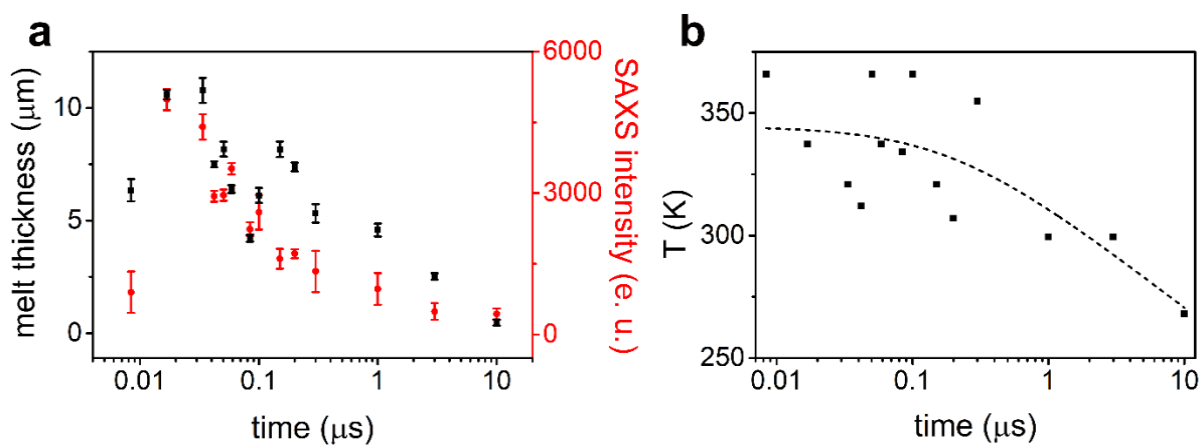

**Fig. 2.** (a) The time-dependent change of the thickness of the liquid domain (black square) and the integrated intensity of the SAXS region of the difference scattering pattern (red circle). (b) The estimated temperatures of liquid water at each time delay. The dashed lines are shown to guide the eye.

Given that WAXS is only sensitive to the local structures at short distances, the difference intensity of the diffuse scattering at the WAXS region is directly proportional to the total amount of liquid water in the probed volume at a specific time delay. Thus, by comparing the integrated intensity of the diffuse scattering in the WAXS region with that measured from a sample with a known thickness at various temperatures, ranging from 254 K to 366 K[30], we can directly estimate the total thickness of the liquid water in the probed volume at each delay. We performed the fitting analysis by using the liquid water scattering patterns at various temperatures and adjusting the scaling factor. The temperature that gives the best match was chosen as the temperature of the molten domain at each time delay. By using this scaling factor, we converted the integrated WAXS intensity of the experimental data into the thickness of the liquid water. The time-dependent change of the thickness of the molten fraction is shown in Fig 2(a). The amount of liquid water increases at early delays due to the ultrafast melting and reaches a maximum at around 20 ns. However, at later delays, recrystallization occurs due to cooling, causing the proportion of liquid water to decrease to almost zero at 10 μs.

The timescale of the melting process is consistent with a previous study using time-resolved IR spectroscopy[15]. In that study, the time-dependent change of $v_2+L_2$ absorption bands of liquid water reached its maximum at around 25 ns, and a melting time constant of 4.1 ns was reported. However, the timescale of recrystallization (~1 μs) in our present study is faster than those previously reported using time-resolved Mie scattering spectroscopy (within a few ms)[16]. Since the crystallization is limited by the timescale of the heat dissipation to the surrounding in this case, we consider the discrepancy to originate in the current study from using a much lower base temperature (170 K vs. 270 K).

We clearly observe partial melting, which is qualitatively consistent with the results reported as 'superheating' in the previous spectroscopic studies[12-15]. We determined the molten fraction of the sample within the probed volume to be of about 13 % (from the ratio of the thickness, ~11 μm/84 μm), which is lower than that reported of a previous study using time-resolved IR spectroscopy (32.4 %)[15]. Once more, we consider this discrepancy to be caused by our lower base temperature (170 K vs. 270 K).

It is well known that the shape of the scattering pattern of liquid water at WAXS region varies depending on the temperature. Thus, as explained above, by comparing the shape of the diffuse scattering generated by the ultrafast melting with that of liquid water measured at various temperatures, the temperature of the molten fraction can be estimated. The comparisons are shown in Fig. 1(d) and the estimated temperatures of liquid water at each time delay are summarized in Fig. 2(b). The temperature rises from the base temperature (170 K) to ~350 K within our experimental time resolution and then decreases to the melting temperature (~270 K) within 10 μs. The timescale of the heat dissipation is simulated to be about 1 μs by using the thermal diffusivity of ice[31] and is qualitatively in good agreement with the value estimated from Fig 2(b). We note that there can exist a temperature gradient within the sample, thus, the temperature estimated here corresponds to an average value. Consequently, recrystallization can occur at the edge of the pumped volume already when the average temperature is higher than the melting temperature.

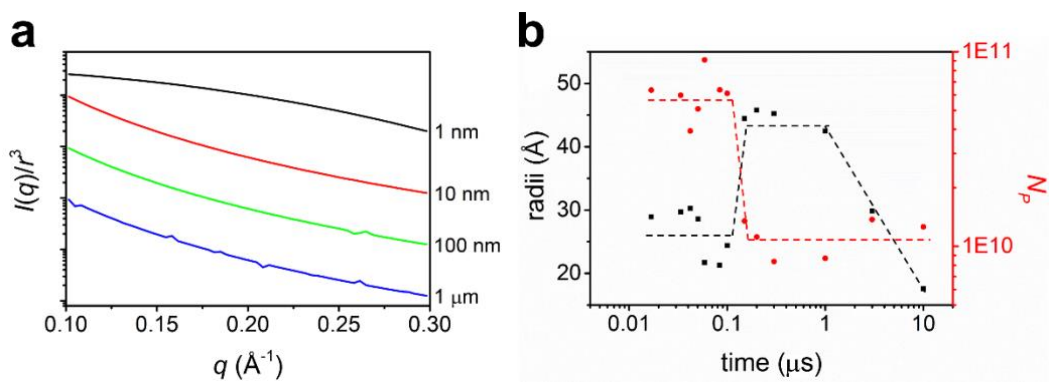

**Fig. 3.** (a) Simulated SAXS intensities of liquid water domains of different radius, normalized by $r^3$. (b) The time dependent changes of the radii of the liquid water domain (black square) and the number of the liquid water domains (red circle) estimated from the SAXS pattern are shown together. The dashed lines are shown to guide the eye.

As shown in Fig 2(a), the integrated SAXS intensity decays much faster than that of WAXS. Since the WAXS intensity is proportional to the total amount of liquid water, the faster decay constant of the SAXS intensity indicates that the SAXS intensity decreases even when the total amount of molten fraction remains stable. To understand this behavior, we simulated the SAXS intensity of differently sized molten domains. If we consider that the total amount of molten fraction in each simulation is the same (by normalizing by $r^3$), the SAXS intensity within our $q$ range decreases with increasing the size of the molten domain as shown in Fig. 3(a). Thus, our results (a faster SAXS intensity decay than that of the WAXS) can be well explained by the increase of the size of the molten domains, but decrease of their total number. We interpret this observation as coalescence of the adjacent domains at around 100 ns.

By analyzing the SAXS patterns, taking into account the total amount of liquid water, as estimated from the WAXS data, the size and the number of liquid domains at each delay can be determined. If we assume molten domains as homogeneously distributed spherical particles in the ice medium, the experimental SAXS patterns at each time delay can be fitted with the theoretical SAXS pattern calculated by the following equation (See Method section for details):

$$\Delta I(q) = cN_p \left(\Delta\rho_{electron} \frac{4\pi}{3} r^3\right)^2 \left[3\frac{sin(rq)-rqcos(rq)}{(rq)^3}\right]^2 \quad (1)$$

where $c$ is a scaling factor between the theory and the experiment, $N_p$ is the number of molten domains, $\Delta\rho_{electron}$ is the difference in the number density of the electrons between the liquid and ice phases of water, $r$ is the radius of the liquid domain, and $q$ is the momentum transfer. We note that the only free parameter that was adjusted during the fitting analysis is the radius of the liquid domain, $r$. The number of the molten domain, $N_p$ is dependent on the radius ($r$) and calculated by the following equation.

$$N_p(t) = \frac{V_{water}}{V_p} = \frac{V_{water}}{\frac{4\pi}{3}r^3} \quad (2)$$

$V_{water}$ is the total volume of liquid water within the probed volume generated by the ultrafast melting of ice which was estimated from the WAXS analysis (Fig. 2(a)). $V_p$ is the volume of each individual domain and is calculated by using the radius ($r$).

The theoretical SAXS patterns obtained from the fitting analysis are shown together with the experimental data in Fig. 1(c). The estimated radius ($r$) and the calculated number ($N_p$) of the molten domains as a function of time delays are summarized in Fig. 3(b). Initially, liquid domains with a radius of 2.5 nm are formed within 16.8 ns and remain for ~100 ns. At around 100–200 ns, the radius increases to *ca*. 4.5 nm and the number of domains decreases by a factor of 6. The previous is consistent with our proposed coalescence scenario, inferred from the integrated SAXS intensity and the difference in decay times between SAXS and WAXS.

After about 1 μs, the radius decreases again to less than 2 nm, while the number of domains stays the same, corresponding to the recrystallization of the molten fraction. We note that the timescale of the recrystallization is consistent with the cooling process estimated from the WAXS analysis, as shown in Fig 2(b).

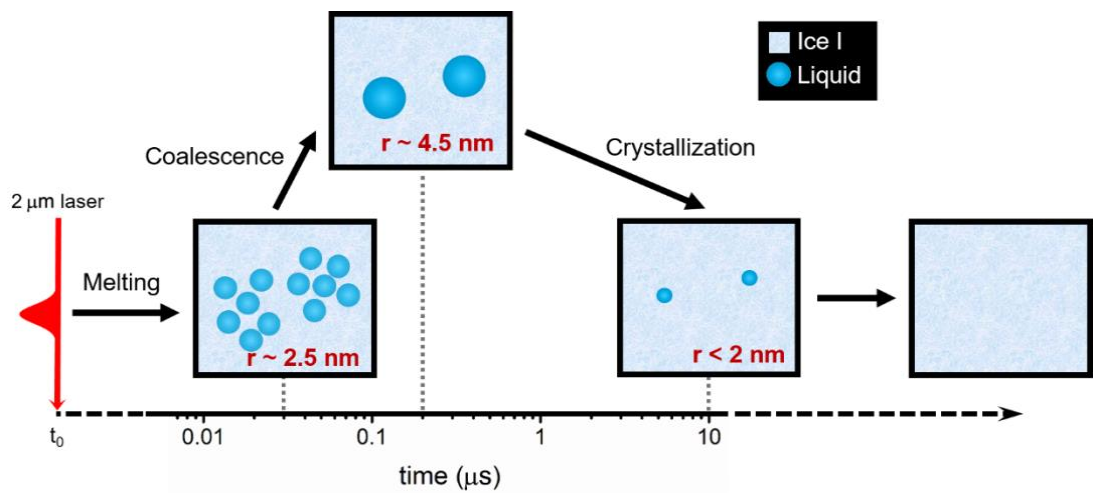

**Fig. 4.** Schematic picture of a proposed model of the melting and recrystallization dynamics of ice I. After the IR irradiation, ice is partially melted by the rapid heating. Initially, the radius of the molten domains is around 2.5 nm. After ~100 ns, the liquid domains grow to a radius of 4.5 nm by coalescence of the adjacent domains. The recrystallization of the liquid domains occurs at ~10 μs.

In Fig. 4, we summarize the model of the melting and recrystallization dynamics of ice I based on the conclusion obtained, by combining the results of SAXS and WAXS analysis. After the IR irradiation, ice is only partially melted by the rapid heating, even if the temperature of the liquid water increased to ~350 K, in agreement with the superheating of ice. The molten fraction reaches its maximum at around 20 ns and the average radius of the liquid domains at this stage is of around 2.5 nm. At around ~100 ns, the liquid domains grow to a radius of 4.5 nm through the coalescence of approximately six adjacent domains. As the temperature of the water in the probed volume cools down by heat dissipation, the recrystallization of the liquid domain occurs in a broad time range up to several μs, and the size of the liquid domain decreases continuously. After approximately 10 μs, the sample is completely frozen. As we have previously noted, the timescale of the recrystallization in the current study is much faster than that from a previous study[16], but it is consistent with the recrystallization rate increasing at lower base temperatures.

In summary, we have performed a time-resolved x-ray scattering experiment on the effects of IR-irradiation-induced ultrafast heating of crystalline ice and have elucidated the melting and recrystallization dynamics of crystalline ice with direct structural information. Our study demonstrates that by combining the structural sensitivity of x-ray scattering experiments on different length scales with ultrafast temporal resolution, a comprehensive picture on the complicated processes such as ultrafast melting and recrystallization of ice can be elucidated. Finally, we note that by utilizing the experimental and analysis protocol established here, future experiments focused on the femtosecond time regime can reveal important structural details in the ice phase of the initial melting process, such as the mechanism of the formation of superheated water or the energy dissipation through the ice lattice.

**Methods**

**Preparation of ice I sample**

Ice I samples were prepared in a 0.1 mm-thick Cu-grid that has a total diameter of 10 mm and contains holes of 0.8 mm or 1.5 mm in diameter. The Cu-grid was first filled with ultra-pure deionized water and then dipped into liquid nitrogen to form ice I.

**Time-resolved x-ray scattering experiment**

An x-ray scattering experiment with a pump-probe scheme was performed at XSS-FXS beamline of PAL-XFEL[28,29]. Ice I samples on a Cu-grid were loaded on a custom-made sample holder connected to a cryostat (ST-400, Janis Research Company). The temperature of the cryostat was controlled by a silicon diode and a cryogenic temperature controller (DT-670 and Model 335, respectively, Lake Shore Cryotronics). The sample was placed inside a vacuum chamber with a pressure of less than 0.1 Pa and cooled with liquid nitrogen. The base temperature of the sample during the measurement was maintained at 170 K. Femtosecond optical laser pulses at 2 μm wavelength with an energy of 250 μJ/pulse were generated from an optical parametric amplifier, which was pumped by pulses at 800 nm center wavelength of a Ti:sapphire regenerative amplifier. The laser beam was focused to a spot of 70 μm (FWHM) diameter. Femtosecond x-ray pulses were generated from the x-ray free-electron laser (XFEL) at PAL-XFEL by self-amplified spontaneous emission (SASE). The x-ray beam has a mean energy of 9.7 keV and an energy bandwidth of 0.3 % ($\Delta E/E$) and was focused to a spot of 19 μm × 32 μm. An optical laser pulse was spatially overlapped with an x-ray pulse with an angle of 20°. The $q$ range of 0.1–3.2 Å$^{-1}$ was covered by a large area CCD detector (MX225-HS, Rayonix) at a distance of 250 mm. An ultrafast temperature jump (T-jump) within the sample was induced by an optical laser pulse (pump) and the time-dependent change of the sample was probed by an x-ray pulse (probe). To probe a fresh sample for each measurement, the sample holder was moved to a fresh position prior to each pump-probe measurement. Laser-off images (probe-only) prior to the pump-probe measurement were acquired as references for each position and used for obtaining the difference x-ray scattering pattern at each time delay. The scattering images were acquired at the following time delays: -8.4 ns, 8.4 ns, 16.8 ns, 33.6 ns, 42 ns, 50.4 ns, 58.8 ns, 84 ns, 100 ns, 150 ns, 200 ns, 300 ns, 1 μs, 3 μs, and 10 μs.

**Integrated intensities of the difference scattering pattern**

As a measure of time-dependent changes of the difference scattering pattern associated with the melting and recrystallization dynamics of ice, the integrated intensities at SAXS and WAXS regions were used. The $q$ range from 0.10 Å$^{-1}$ to 0.30 Å$^{-1}$ was used for the SAXS region and from 1.00 Å$^{-1}$ to 2.73 Å$^{-1}$ was used for the WAXS region. The $q$ range where the Bragg peaks of ice I appear was excluded. The integrated intensities are shown in Fig. 2(a).

**Temperature and thickness estimation of the molten domain**

We estimated the temperature of the molten domain by comparing the shape of the diffuse scattering at WAXS region with that of liquid water measured at various temperatures, ranging from 254 K to 366 K [30]. We performed the fitting anlysis by adjusting the scaling factor, and the temperature that gives the best match was chosen as the temperature of the molten domain at each time delay. The comparisons at all time delays are shown in Fig. S3 and the time-dependent change of the temperature is shown in Fig. 2(b). From the fitting analysis, we obtained the optimal scaling factor between our experimental diffuse scattering and the normalized scattering pattern of liquid water[30]. By using this scaling factor, we converted the integrated WAXS intensity of our experimental data into the thickness of the liquid water. The time-dependent change of the estimated thicknesses of the molten domain is shown in Fig. 2(a).

**Radius estimation of the molten domain**

By analyzing the difference SAXS intensities, the size and the number of the liquid water domains at each time delay were determined. Equation S1 describes the SAXS pattern of a spherical particle with a radius of $r$[32].

$$\Delta I(q)_{single} = \left(\Delta\rho_{electron}\frac{4\pi}{3}r^3\right)^2 \left[3\frac{sin(rq) - rqcos(rq)}{(rq)^3}\right]^2 \quad (S1)$$

If we assume molten domains as spherical particles in the ice medium, the experimental SAXS patterns at each time delay can be fitted with the theoretical SAXS pattern calculated by the following equation.

$$\Delta I(q) = cN_p \left(\Delta\rho_{electron}\frac{4\pi}{3}r^3\right)^2 \left[3\frac{sin(rq) - rqcos(rq)}{(rq)^3}\right]^2 \quad (S2)$$

where $c$ is a scaling factor between the theory and the experiment, $N_p$ is the number of the molten domain, $\Delta\rho_{electron}$ is the difference in the number density of the electron between the liquid and ice phases of water, $r$ is the radius of the liquid domain, and $q$ is the momentum transfer. The free fitting parameter that was adjusted during the fitting analysis is the radius of the liquid domain, $r$. The electron densities of liquid water and ice are calculated from the mass densities of liquid water and ice, and they are 0.98 and 0.93 g/cm$^3$, respectively. Since the mass density varies slightly depending on the temperature, the average value was used. A scaling factor between the theory and the experiment, $c$, was determined from the WAXS analysis by comparing our experimental diffuse scattering with the normalized scattering pattern of liquid water [30] (See *Temperature and thickness estimation of the molten domain* section for details). The number of the molten domains, $N_p$, was calculated by Equation S3.

$$N_p(t) = \frac{V_{water}}{V_p} = \frac{V_{water}}{\frac{4\pi}{3}r^3} \quad (S3)$$

$V_p$ is a volume of a single domain and is calculated from the radius $r$. $V_{water}$ is the total volume of liquid water within the probed volume and was estimated from the WAXS analysis.

In a more realistic picture, the molten domains with slightly different sizes can exist at the same time with a certain distribution. In order to take this into account, we used a normal distribution of the radius $f(r)$ for calculating the theoretical SAXS patterns instead of using a single value of $r$.

$$f(r) = \frac{e^{-\frac{1}{2}\left(\frac{r - r_{mean}}{\sigma_r}\right)^2}}{\sigma_r\sqrt{2\pi}} \quad (S4)$$

where $\sigma_r$ is the variance and $r_{mean}$ is the mean value of the radius. For the representative fitting results, the ratio between the variance and the mean value of the radius, ($\sigma_r/r_{mean}$), was set to be 0.2. We note that the distribution of radius effectively reduce the oscillatory features in the theoretical SAXS pattern by averaging trigonometric functions with different period and the choice of the degree of the distribution does not affect our conclusion. The theoretical SAXS patterns obtained from the fitting analysis are shown together with the experimental data in Fig.

S2. The estimated radius ($r$) and the calculated number ($N_p$) of the molten domains as a function of time delays are shown in Fig. 3(b).

**Estimation of the heat diffusion**

After the ultrafast T-jump, the cooling process occurs due the the heat dissipation. The timescale of the cooling process is estimated by using the thermal diffusivity. The initial temperature of the irradiated area right after the ultrafast T-jump was estimated to be ~350 K based on the WAXS analysis as shown in Fig. 2(b), whilst the area that is not irradiated has the base temperature of 170 K. The thermal diffusivity of ice I varies from $2.87 \times 10^{-6}$ m$^2$/s at 170 K to $1.17 \times 10^{-6}$ m$^2$/s at 273 K[31]. The time-depenent changes of the temperature near the center (red) and the edge (green) of the irradiated area were calculated and are shown in Fig. S5. Due to the differences in the timescale of the cooling, a temperature gradient within the sample can be generated and the recrystallization can occur at the edge of the pumped volume already when the average temperature is higher than the melting temperature. The simulated cooling time scale is in qualitative agreement with that of our experiment shown in Fig 2(b). We note that a temperature gradient along the beam axis may also exist since the front of the sample absorbs more energy than the back and the timescale of the heat dissipation can be faster than the values shown in Fig. S5 that is calculated from the simplified 1-dimensional case.

**Acknowledgment**

This work was supported by the National Research Foundation of Korea (NRF) grant funded by the Korea government (MSIT) (No. 2019R1C1C1006643 and 2020R1A5A1019141). The experiments were performed at beamline XSS of PAL-XFEL (proposal 2019-1st-XSS-008) funded by the Korea government (MSIT). This work also has been supported by a European Research Council Advanced Grant under project 667205 and the Swedish National Research Council.

**Author Contributions**

K.H.K. designed and supervised the study. K.A.-W. designed sample preparation. K.A.-W. and A.S. designed sample holder. K.A.-W., M.L.-P., A.S., and T.E. prepared ice samples. K.H.K., K.A.-W., A.N., A.S., F.P., and H.P. designed experimental setup, chamber, and laser geometry. K.H.K., K.A.-W., A.N., A.S., F.P., H.P., M.L.-P., C.Y., T.E., T.J.L., S.Y., S.J., J.H.L., I.E., and M.K. performed the x-ray experiments. C.Y., K.H.K., K.N., S.Y., and S.J. analyzed the data. C.Y., K.H.K., A.N., M.L.-P., and F.P. wrote the manuscript.

# Supplemental Material for "Melting Domain Size and Recrystallization Dynamics of Ice Revealed by the Time-Resolved X-ray Scattering"

Cheolhee Yang[1], Marjorie Ladd-Parada[2], Kyeongmin Nam[1], Sangmin Jeong[1], Seonju You[1], Alexander Späh[2], Harshad Pathak[2], Tobias Eklund[2], Thomas J. Lane[3], Jae Hyuk Lee[4], Intae Eom[4], Minseok Kim[4], Katrin Amann-Winkel[2], Fivos Perakis[2], Anders Nilsson[2], and Kyung Hwan Kim[1,*]

[1]Department of Chemistry, Pohang University of Science and Technology (POSTECH), Pohang, Gyeongbuk 37673, Republic of Korea.
[2]Department of Physics, AlbaNova University Center, Stockholm University, SE-106 91 Stockholm, Sweden.
[3]SLAC National Accelerator Laboratory, 2575 Sand Hill Road, 'Menlo Park, CA 94025, USA.
[4]Pohang Accelerator Laboratory, Pohang, Gyeongbuk 37673, Republic of Korea.
*Corresponding author. Email: kyunghwankim@postech.ac.kr

# Supplementary Figures

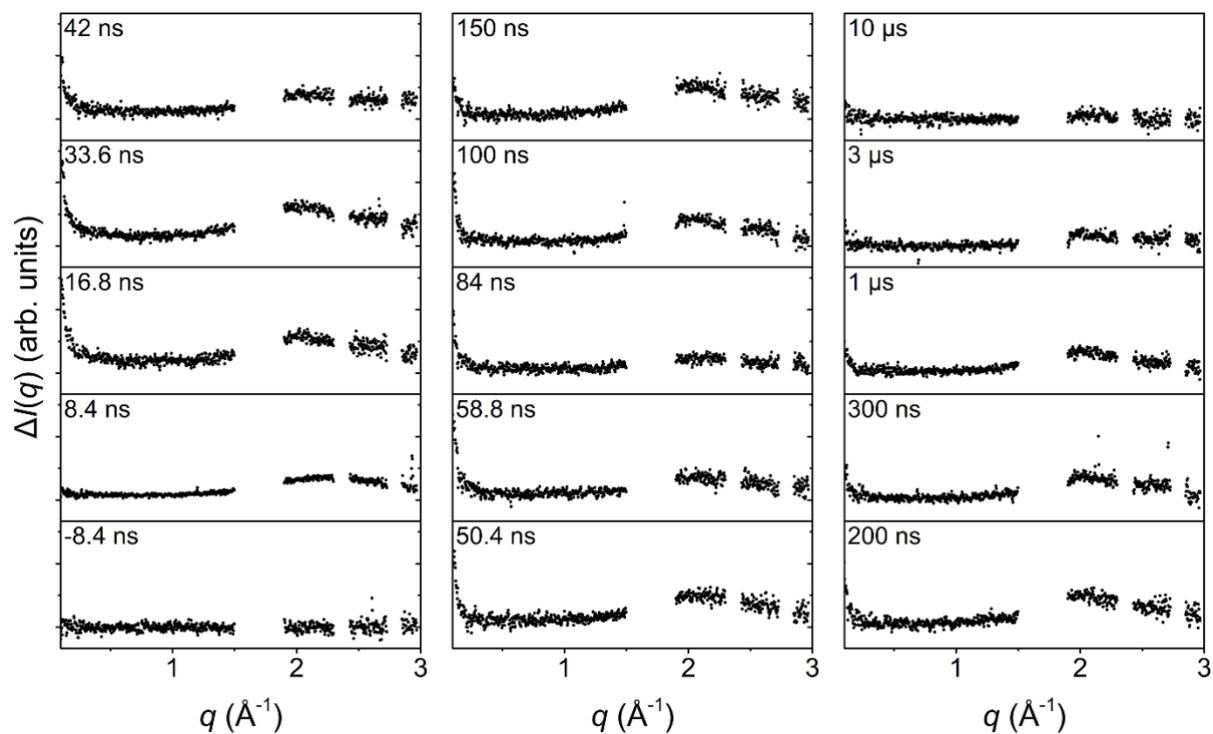

**Fig. S1.** The difference scattering intensities, $\Delta I(q)$, measured at various time delays from -8.4 ns to 10 μs. The regions nearby the Bragg diffraction peaks are not shown for clarity.

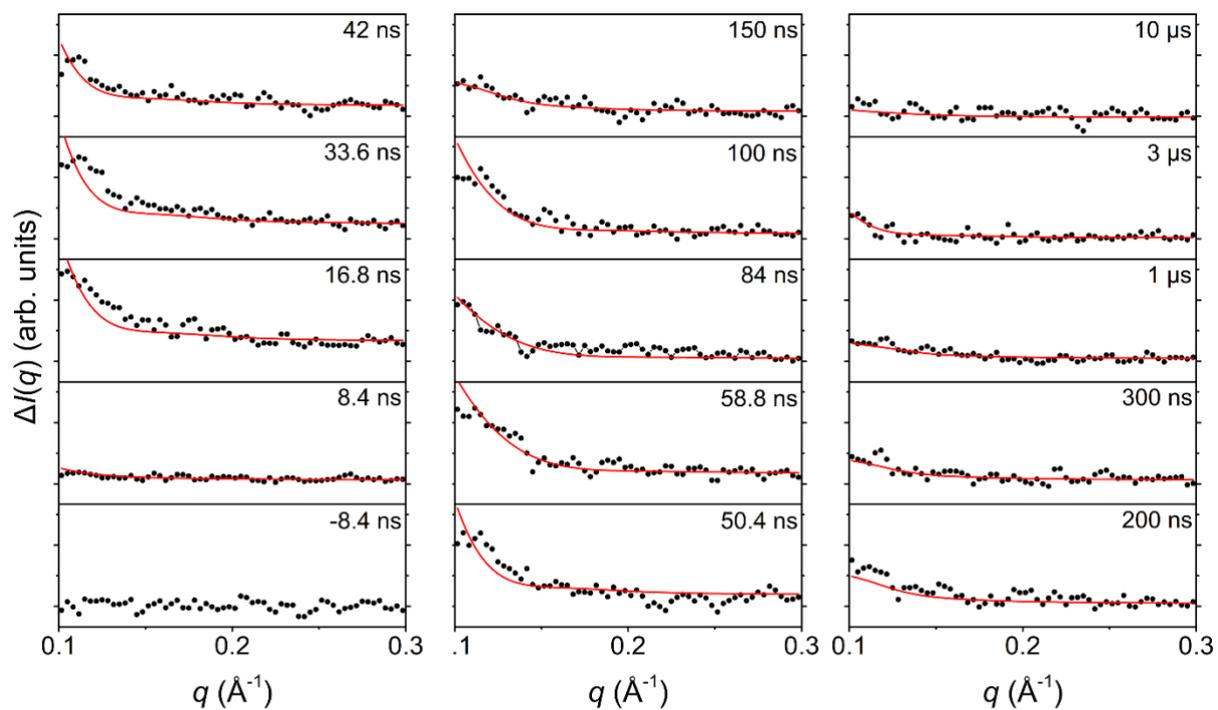

**Fig. S2.** The difference scattering intensities, $\Delta I(q)$, in the SAXS region at time delays from -8.4 ns to 10 μs (black) are shown together with the fitting results (red line).

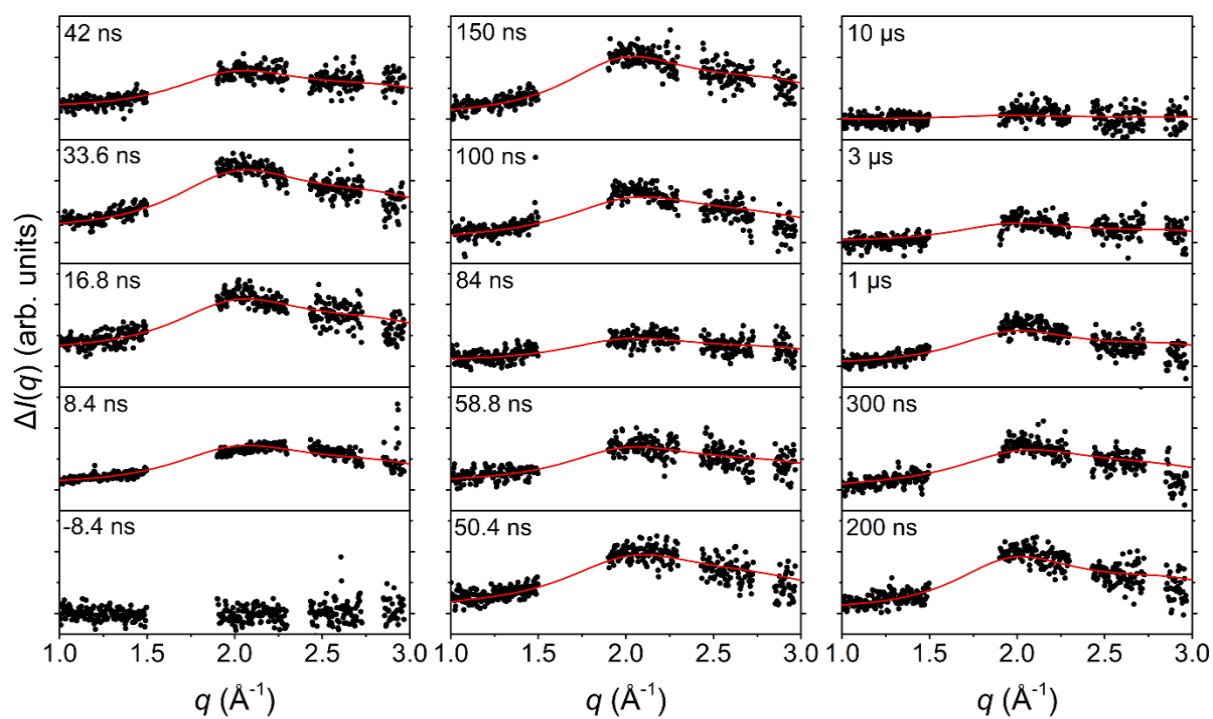

**Fig. S3.** The difference scattering intensities, $\Delta I(q)$, in the WAXS region at time delays from -8.4 ns to 10 μs (black circle) are shown together with their best match with the the scaled scattering pattern of liquid water from the reference among various temperatures from 254 K to 366 K[1]. The regions nearby the Bragg diffraction peaks are not shown for clarity.

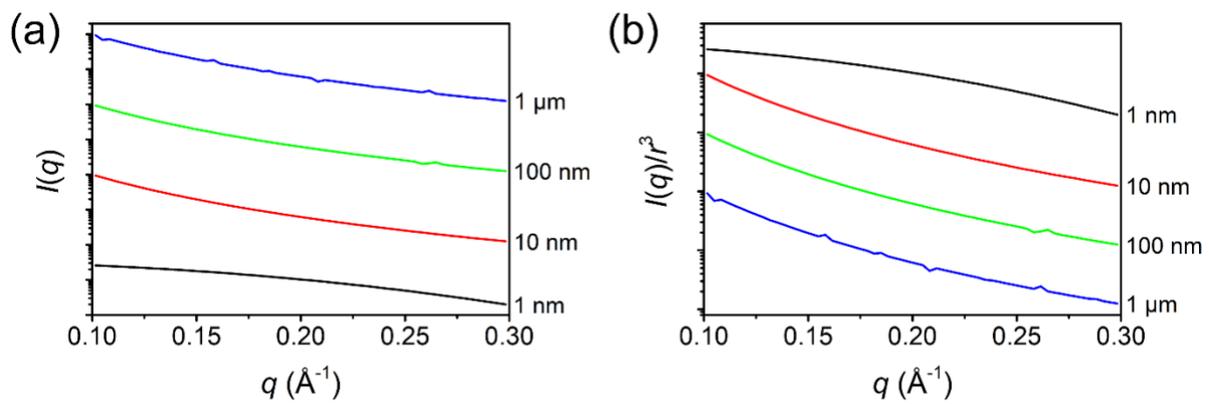

**Fig. S4.** Simulated SAXS intensities of liquid water domains of different radius (a) before and (b) after the normalization by $r^3$.

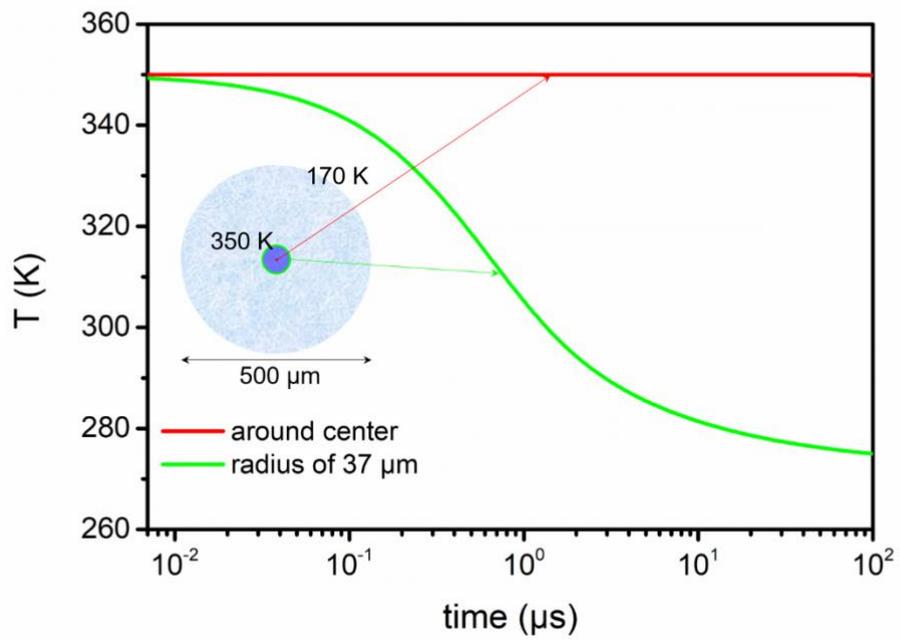

**Fig. S5.** Estimated timescales for heat dissipation in the sample. The time-depenent changes of the temperature near the center (red) and edge (green) of the irradiated area are shown.

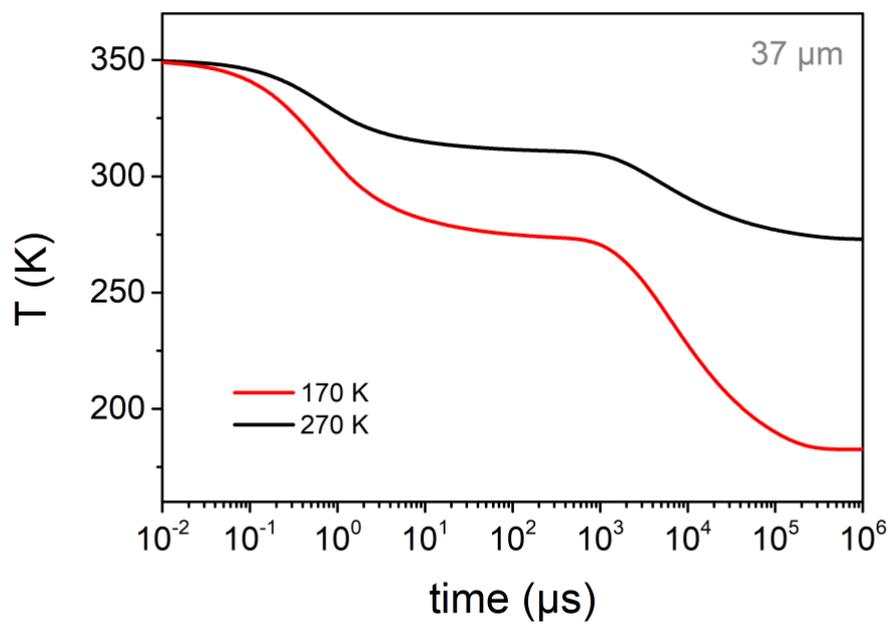

**Fig. S6.** Estimated timescales for heat dissipation in the sample with the base temperature of 270 K (black) and 170 K (red) near the edge of the irradiated area.